\documentclass[letterpaper,journal,onecolumn]{IEEEtran}

\usepackage{amssymb}
\usepackage{amsmath}
\usepackage{amsfonts}
\usepackage{epsfig}
\usepackage{graphicx}
\usepackage{color}
\usepackage{crop}
\usepackage{psfrag}
\usepackage{subfig}
\usepackage{citesort}
\usepackage{cite}

\begin{document}

\title{{A note on the bivariate distribution \\representation of two perfectly correlated \\random variables by Dirac's $\delta$-function}}

\author{Andr\'{e}s Alay\'{o}n Glazunov$^{\rm a}$$^{\ast}$\thanks{$^\ast$Corresponding author. Email: aag@ee.kth.se
\vspace{6pt}} and Jie Zhang$^{\rm b}$\\
$^{\rm a}${KTH Royal Institute of Technology, Electrical Engineering,} \\{Teknikringen 33, SE-100 44 Stockholm, Sweden}; \\$^{\rm b}${University of Sheffield, Electronic and Electrical Engineering,} \\{Mappin Street, Sheffield, S1 3JD, UK }\\\vspace{6pt}}

\maketitle

\begin{abstract}
In this paper we discuss the representation of the joint probability density function of perfectly correlated continuous random variables, i.e., with correlation coefficients $\rho\!\!=\!\!\pm1$, by Dirac's $\delta$-function. We also show how this representation allows to define Dirac's $\delta$-function as the ratio between bivariate distributions and the marginal distribution in the limit $\rho\rightarrow \pm1$, whenever this limit exists. We illustrate this with the example of the bivariate Rice distribution.
\end{abstract}


\section{Introduction}
The performance evaluation of wireless communications systems relies on the analysis of the behavior of signals modeled as random variables or random processes that obey prescribed probability distribution laws. The Pearson product-moment correlation coefficient is commonly used to quantify the strength of the relationship between two random variables. For example, in wireless communications, two-antenna diversity systems take advantage of low signal correlation. The diversity gain obtained by combining the received signals at the two antenna branches depends on the amount of correlation between them, e.g., the lower the correlation the higher the diversity gain, \cite{Simon05}. Moreover, when the correlation coefficient is $1$ or $-1$ it is said that the two signals are perfectly correlated, which implies that there is no diversity gain. In this case the joint probability distribution of the signals is not well-defined and, to the best knowledge of the authors, a thorough discussion of these limit cases is not found in the literature. Here, we fill this gap.

We therefore investigate in this letter the representation of the joint probability distribution in the limit case of perfectly correlated random variables. And show that this is achieved by Dirac's $\delta$-function, \cite{Arfken+Weber2001}. Based on this, we propose a representation of Dirac's $\delta$-function as the ratio between bivariate distributions and the marginal distribution in the limit $\rho\rightarrow \pm1$. An example of the evaluation of the limit is provided for the bivariate Rice distribution.

\section{Dirac's $\delta$-function representation}
Let $X\in\mathbb{R}$ and $Y\in\mathbb{R}$ be two random variables defined on some probability space. Thus, the Pearson
product-moment correlation coefficient between them is defined as, \cite{Papoulis1991}
\begin{equation}\label{Eq:1}
\rho=\frac{E[(x-\mu_X)(y-\mu_Y)]}{\sigma_X\sigma_Y},
\end{equation}
where $E[.]$ denotes the expected value operator, $\mu_X=E[x]$ and $\mu_Y=E[x]$ denote the mean values, and $\sigma_X$ and $\sigma_Y$ denote the standard deviations, i.e.
\begin{equation}\label{Eq:2}
\sigma_{X}=\sqrt{E[x^2]-{\mu_X}^2} \text{~and~} \sigma_{Y}=\sqrt{E[y^2]-{\mu_Y}^2}.
\end{equation}
Definition \eqref{Eq:1} ensures that $-1\leq\rho\leq1$.

Let $f_{X,Y}(x,y)$ denote the \emph{joint} probability density function\footnote{Throughout the paper we shorten probability density function to pdf or just probability distribution.}of the two random variables $X$ and $Y$. By definition we have that
\begin{equation}\label{Eq:3}
f_{X,Y}(x,y) = f_{Y\mid X}(y\mid x)f_{X}(x)=f_{X\mid Y}(x\mid y)f_{Y}(y),
\end{equation}
where $f_{Y\mid X}(y\mid x)$ and $f_{X\mid Y}(x\mid y)$ are the \emph{conditional} distributions, $f_{X}(x)$ and $f_{Y}(y)$ denote the \emph{marginal} distributions of variables $X$ and $Y$, respectively.

Let's further assume that random variables $X$ and $Y$ are identically distributed, with mean values of equal or opposite signs. We state then that the joint probability distribution $f_{X,Y}(x,y)$ can be compactly represented as follows
\begin{equation}\label{Eq:4}
f_{X,Y}(x,y)= \left\{
\begin{array}{cll}
f_{X}(x)\delta(x\mp y)&\text{if }\rho = \pm1&\text{and }\mu_{X}=\pm\mu_{Y},\\
0&\text{if }\rho = \pm1&\text{and }\mu_{X}=\mp\mu_{Y}\neq 0,
\end{array} \right.
\end{equation}
where the upper and the lower signs should be evaluated separately, $\delta(.)$ is the $\delta$-function or Dirac's $\delta$-function and $f_{X}(x)\delta(x\mp y)=f_{Y}(y)\delta(y\mp x)$.

Indeed, for perfectly correlated random variables $\rho=\pm1$ we have from \eqref{Eq:1} that
\begin{equation}\label{Eq:5}
\pm\sigma^{2}=E[xy]\mp\mu^2,
\end{equation}
where the $(+)$ and $(-)$ signs on the left hand side of \eqref{Eq:4} correspond to $\rho=1$ and $\rho=-1$, respectively. On the righthand side the $(+)$ and $(-)$ signs correspond to opposite sign mean values ($\mu=\mu_{X}=-\mu_{Y}$) and equal sign mean values ($\mu=\mu_{X}=\mu_{Y}$), respectively.

From \eqref{Eq:2} we have that
\begin{equation}\label{Eq:6}
\sigma^2=E[x^2]-\mu^2.
\end{equation}
Consider now $E[(x-ky)^2]=0$. Thus, solving for $k$ and combining \eqref{Eq:5} and \eqref{Eq:6} gives
\begin{subequations}\label{Eq:7}
\begin{eqnarray}
k &=& \frac{E[x y]\pm\sqrt{E^2[x y]-E^2[x^2]}}{E[x^2]}  \\ \label{Eq:7a}
  &=& \frac{\pm\sigma^2\pm\mu^2\pm\sqrt{(\pm\sigma^2\pm\mu^2)^2-(\sigma^2+\mu^2)^2}}{\sigma^2+\mu^2}\label{Eq:7b},
\end{eqnarray}
\end{subequations}
where we have used our assumption of identically distributed variables.
A straightforward analysis of \eqref{Eq:7b} leads to three possibilities: (I), $k=1$ if $\rho=1$ and $\mu=\mu_{X}=\mu_{Y}=0$ or $\mu=\mu_{X}=\mu_{Y}\neq 0$, (II), $k=-1$ if $\rho=-1$ and $\mu=\mu_{X}=\mu_{Y}=0$ or $\mu=\mu_{X}=-\mu_{Y}\neq 0$, and (III), there is no (real) solution if $\rho=1$ and $\mu=\mu_{X}=-\mu_{Y}\neq 0$ or if $\rho=-1$ and $\mu=\mu_{X}=\mu_{Y}\neq 0$.
Hence, summarizing, we find that
\begin{equation}\label{Eq:8}
k= \left\{
\begin{array}{rl}
\pm 1 & \text{if } \rho=\pm 1 \text{ and } \mu=\mu_{X}=\pm\mu_{Y},\\
\{\emptyset\} & \text{if } \rho= \pm 1 \text{ and } \mu=\mu_{X}=\mp\mu_{Y}\neq 0.\\
\end{array} \right.
\end{equation}
Clearly, $k=\pm1$ in the first case implies a linear, deterministic relationship between random variables $X$ and $Y$ with probability one. Thus, we can write
\begin{equation}\label{Eq:9}
f_{Y\mid X}(y\mid x)=\delta (y\mp x) \text{ and } f_{X\mid Y}(x\mid y)=\delta (x\mp y),
\end{equation}
since $\delta (y\pm x)=\delta (x\pm y)$.
Now, inserting \eqref{Eq:8} into \eqref{Eq:9} gives
\begin{equation}\label{Eq:10}
f_{X,Y}(x,y)=f_{X}(x)\delta (y\mp x)=f_{Y}(y)\delta (x\mp y).
\end{equation}
On the other hand, $k=\{\emptyset\}$ implies in the second case that there exist no linear, deterministic relationship between random variables $X$ and $Y$ and therefore a perfect correlation between the variables cannot be observed. Hence, in this case
\begin{equation}\label{Eq:11}
f_{X,Y}(x,y)=0.
\end{equation}
This ends our derivation.

Let us consider bivariate distributions of identically distributed random variables (with mean values of equal or opposite signs) as previously. We further assume they also depend on the correlation coefficient $\rho$ as a parameter, and are well-defined functions for $0 \leq |\rho| < 1$. However, such distributions are in general not well-defined for $|\rho| = 1$ since terms of the form $(1\pm\rho)^{-1}$, $(1\pm\rho)^{-1/2}$, $(1\pm\rho^2)^{-1}$ and/or $(1\pm\rho^2)^{-1/2}$ may appear in their arguments. Examples of such bivariate distributions are the Gaussian (for both complex and real variables) and the family of distributions derived from the complex Gaussian multivariate distribution , \cite{Papoulis1991}, such as the bivariate Rice , \cite{Abu-Dayya1994}, the bivariate Nakagami, \cite{Nakagami1960}, and the bivariate Weibull distributions , \cite{J:Sagias_J4_2005}. Henceforth, we denote such a bivariate distribution as $f_{X,Y}(x,y;\rho)$ to make a distinction between the joint distribution and bivariate distribution, i.e., we explicitly point out the dependence on $\rho$ for $0 \leq |\rho| < 1$. Therefore, we can write
\begin{equation}\label{Eq:12}
f_{X,Y}(x,y) = f_{X,Y}(x,y;\rho) \text{ if } 0 \leq |\rho| < 1.
\end{equation}
However, as we have shown above the singularities can be represented through the $\delta$-function when $\rho=1$ and $\mu_{X}=\mu_{Y}$, or $\rho=-1$ and $\mu_{X}=-\mu_{Y}$ or $\rho=\pm1$ and $\mu_{X}=\mu_{Y}=0$. Thus
\begin{equation}\label{Eq:13}
f_{X,Y}(x,y)=\lim_{\rho\rightarrow\pm1} f_{X,Y}(x,y;\rho) = f_{X}(x)\delta(x\mp y).
\end{equation}
Hence, Dirac's $\delta$-function has the following representation
\begin{equation}\label{Eq:14}
\delta(x\mp y)=\lim_{\rho\rightarrow\pm1} \frac{f_{X,Y}(x,y;\rho)}{f_{X}(\pm y)}.
\end{equation}

Assume now that $X_{c}\in\mathbb{C}$ and $Y_{c}\in\mathbb{C}$ are two random variables defined on some probability space. Thus, the Pearson
product-moment correlation coefficient between them is defined as, \cite{Papoulis1991}
\begin{equation}\label{Eq:14a}
\rho_{c}=\frac{E[(x_{c}-\mu_{X_{c}})(y_{c}^{*}-\mu^{*}_{Y_{c}})]}{\sigma_{X_{c}}\sigma_{Y_{c}}},
\end{equation}
where $E[.]$ denotes the expected value operator, $(.)^{*}$ denotes complex-conjugate, $\mu_{X_{c}}=E[x_{c}]$ and $\mu_{Y_{c}}=E[x_{c}]$ denote the mean values, and $\sigma_{X_{c}}$ and $\sigma_{Y_{c}}$ denote the standard deviations, i.e.
\begin{equation}\label{Eq:14b}
\sigma_{X_{c}}=\sqrt{E[|x_{c}-\mu_{X_{c}}|^2]} \text{~and~} \sigma_{Y_{c}}=\sqrt{E[|y_{c}-\mu_{Y_{c}}|^2]}.
\end{equation}
We further assume that $\rho_{c}=\rho_{c}^{*}$, i.e., $\rho_{c}\in\mathbb{R}$.
In this case if $X_{c}$ and $Y_{c}$ are identically distributed with mean values of equal or opposite signs, then the following holds true
\begin{align}\label{Eq:14c}
&|\rho_{c}|=1 \Rightarrow \rho=\pm1 \text{ if } x=|x_{c}|\text{, }y=\pm|y_{c}|,
\end{align}
where, the upper and the lower signs should be evaluated separately. The converse, i.e., $|\rho|=1 \Rightarrow |\rho_{c}|=1$, doesn't hold in general. This follows from the definition of the correlation coefficients \eqref{Eq:1} and \eqref{Eq:14a}.

\section{An example: the bivariate Rice distribution}
Consider the bivariate Rice distribution of two identically distributed random variables  $X\in\mathbb{R}$ and $Y\in\mathbb{R}$
\begin{align}\label{Eq:15}
& f_{X,Y}(x,y;\rho_{c}) = \frac{(1+K)^2x y}{2\pi\beta^2(1-\rho_{c}^2)}\mathrm{e}^{-\frac{2K}{1+\rho_{c}}-\frac{(1+K)(x^{2}+y^{2})}{2\beta(1-\rho_{c}^2)}}  \\
& \times \int\limits_{0}^{2\pi} \mathrm{e}^{\frac{\rho_{c}(1+K)xy\cos\theta}{\beta(1-\rho_{c}^2)}}I_{0} \Bigg(\sqrt{\frac{2K(1+K)(x^2+y^2+2xy\cos\theta)}{(1+\rho_{c})^2\beta}}\Bigg)\mathrm{d}\theta, \nonumber
\end{align}
where $\beta=\frac{x^2}{2}=\frac{y^2}{2}$ is the average power of, e.g., the signal received at one antenna, $K$ is the power of the line-of-sight(LOS) component to the scattered power and $\rho_{c}$ is the correlation coefficient \eqref{Eq:14a}, and $I_{0}(.)$ is the modified Bessel function of $0$th order.
We see from \eqref{Eq:15} that $f_{X,Y}(x,y;\rho_{c})$ is not well-defined for $\rho_{c}=\pm 1$, so it should be evaluate in the limit $\rho_{c}\rightarrow1$ to obtain a value if it exists.

Evaluating the limit by regrouping terms and using the Algebraic Limit Theorem gives
\begin{align}\label{Eq:16}
& \lim_{\rho_{c}\rightarrow1} f_{X,Y}(x,y;\rho_{c}) = \frac{(1+K)^2xy\mathrm{e}^{-K-\frac{(1+K)xy}{2\beta}}}{2\beta^2}G_{1}(x,y)\\
& \times \int\limits_{0}^{2\pi}G_{2}(x,y,\theta)I_{0}\Bigg(\sqrt{\frac{K(1+K)(x^2+y^2+2xy\cos\theta)}{2\beta}}\Bigg)\mathrm{d}\theta, \nonumber
\end{align}
where
\begin{equation}\label{Eq:17}
G_{1}(x,y)=\lim_{\rho_{c}\rightarrow1}\frac{\mathrm{e}^{-\frac{(1+K)(x-y)^2}{2\beta(1-\rho_{c}^2)}}}{\sqrt{\pi}(1-\rho_{c}^2)^{1/2}},
\end{equation}
\begin{equation}\label{Eq:18}
G_{2}(x,y,\theta)=\lim_{\rho_{c}\rightarrow1}\frac{ \mathrm{e}^{-\frac{\rho_{c} (1+K)xy(1-\cos\theta)}{\beta(1-\rho_{c}^2)}}}{\sqrt{\pi}(1-\rho_{c}^2)^{1/2}},
\end{equation}
The limits in \eqref{Eq:17} and \eqref{Eq:18} are of the form $\frac{0}{0}$, which can be resolved by noticing that Dirac's-$\delta$ can be represented in terms of the limit
\begin{equation}\label{Eq:19}
 \delta(z)=\lim_{\epsilon\rightarrow0}\frac{\mathrm{e}^{-\frac{z^2}{\epsilon^2}}}{\sqrt{\pi}\epsilon}.
\end{equation}
Thus,
\begin{eqnarray}\label{Eq:20}
G_{1}(x,y)=\delta \big(\sqrt{(1+K)/(2\beta)}(x-y)\big),
\end{eqnarray}
and
\begin{eqnarray}\label{Eq:21}
G_{2}(x,y,\theta)=\delta\big(\sqrt{x y(1+K)(1-\cos\theta)/\beta}\big).
\end{eqnarray}
Let's first consider \eqref{Eq:20}, which by applying the following property of Dirac's $\delta$-function
\begin{eqnarray}\label{Eq:22}
\delta(g(x))=\sum_{n}\frac{\delta(x-x_{n})}{\big|\frac{\mathrm{d}g(x_{n})}{\mathrm{d}x}\big|},
\end{eqnarray}
where the $x_{n}$ satisfy $g(x_{n})=0$, is reduced to
\begin{eqnarray}\label{Eq:23}
G_{1}(x,y)=\sqrt{\frac{2\beta}{1+K}}\delta(x-y),
\end{eqnarray}
Similarly, applying \eqref{Eq:22} to \eqref{Eq:21} we identify
\begin{equation}\label{Eq:24}
g(\theta)=\sqrt{((1+K)x y(1-\cos\theta))/\beta},
\end{equation}
with the modulus of the first derivative given by
\begin{equation}\label{Eq:25}
\Bigg|\frac{\mathrm{d}g(\theta)}{\mathrm{d}\theta}\Bigg|=\sqrt{\frac{(1+K)x y}{2\beta}}\Bigg|\cos\frac{\theta}{2}\Bigg|,
\end{equation}
where we have used some standard mathematical identities and operations. Further, it is straightforward to see that the zeros of \eqref{Eq:24} are given by $\theta_{n}=2\pi n$. Thus,
\begin{equation}\label{Eq:26}
G_{2}(x,y,\theta)=\sqrt{\frac{2\beta}{(1+K)x y}}\sum_{n=0}^{\infty}\delta(\theta-2\pi n),
\end{equation}
The shifting property of Dirac's $\delta$-function
\begin{equation}\label{Eq:27}
\int\limits_{\mathcal{D}}f(x)\delta(x-x_{0})\mathrm{d}x=f(x_{0}),
\end{equation}
where $\mathcal{D}$ is the domain of integration and $x_{0}\in\mathcal{D}$. Thus, inserting \eqref{Eq:23} and \eqref{Eq:26} into \eqref{Eq:15} and applying \eqref{Eq:27} we obtain the final result
\begin{align}\label{Eq:28}
& \lim_{\rho_{c}\rightarrow1} f_{X,Y}(x,y;\rho_{c}) = \delta(x-y)\\
&\times\frac{x(1+K)}{\beta}\mathrm{e}^{-K-\frac{(1+K)x^2}{2\beta}}I_{0}\Bigg(x\sqrt{\frac{2K(1+K)}{\beta}}\Bigg). \nonumber
\end{align}

\section{Conclusions}
The Pearson product-moment correlation coefficient $\rho$ plays a fundamental role in the analysis of the performance of diversity systems for wireless communications. We have shown that for bivariate distributions belonging to the Gaussian family, e.g., Rayleigh, Ricean, Nakagami and Weibull, and perfectly correlated random variables they can be represented by Dirac's $\delta$-function if the corresponding linear dependence exists. The presented result can be used in the analytical and numerical analysis of the diversity performance of perfectly correlated diversity branches. In addition, for nearly perfectly correlated results can be obtained by expanding the joint probability distribution around $\rho\approx1$ or $\rho\approx -1$. Finally, we believe that for the sake of completeness the limiting cases corresponding to $\rho=1$ and/or $\rho=-1$ should also be given when providing the joint probability distribution of two random variables.

\section*{Acknowledgments}

This work has been partly supported by the EU FP6~``GAWIND" project and the EU FP7 IAPP ``IAPP@RANPLAN" project.

\end{document}